\documentclass[11pt]{article}
\usepackage{sao1}
\usepackage{graphicx}

\begin{document}

\title{Characterisation of the magnetic fields of the Herbig Be stars HD 200775 and V380 0ri}
\author{
Alecian E.\inst{1} \and  
Wade G.A.\inst{2} \and 
Catala C.\inst{1} \and
Bagnulo S.\inst{3} \and 
Bohm T.\inst{4} \and
Bouret J.-C.\inst{5} \and
Donati J.-F.\inst{4} \and
Folsom C.P.\inst{2} \and
Landstreet J.D.\inst{6} \and
Silvester J.\inst{2}
}
\institute{
%1
Observatoire de Paris, LESIA, 5, place Jules Janssen, F-92195
Meudon Principal CEDEX, France \and
%2
Dept. of Physics, Royal Military College of Canada, 
PO Box 17000, Stn Forces, Kingston, Canada K7K 4B4 \and
%3
European Southern Observatory, Casilla 19001, Santiago 19, Chile \and
%4
Laboratoire d'Astrophysique, Observatoire Midi-PyrИnИes,
14 avenue Edouard Belin, F-31400 Toulouse, France \and
%5
Laboratoire d'Astrophysique de Marseille, Traverse du Siphon, 
BP8-13376 Marseille Cedex 12, France \and
%6
Dept. of Physics \& Astronomy, University of Western Ontario, 
London, Canada N6A 3K7
}

%\date{03.03.05}{12.04.05}
\maketitle 

\begin{abstract}
The origin of the magnetic fields of the chemically peculiar main sequence Ap/Bp stars is still matter of intense debate. The recent discoveries of magnetic fields in Herbig Ae/Be stars using high resolution data obtained with the spectropolarimeter ESPaDOnS at CFHT provide a strong argument in favour of the fossil field hypothesis. Using a simple oblique rotator model of a centered dipole, we fit the Stokes $V$ LSD profiles of two of these magnetic HAeBe stars, HD 200775 and V380 Ori, as well as their variations on timescales from days to months. We find that in both cases the dipole hypothesis is acceptable and we determine the rotation period $P$, the angle between rotation and magnetic axes $\beta$ and the intensity of the magnetic field at pole $B_{\rm P}$.

\end{abstract}

\section{Introduction}

The chemically peculiar Ap/Bp stars are the only main sequence intermediate mass stars hosting detectable magnetic fields. The fossil origin of the magnetic field is currently the favoured hypothesis. This model states that the magnetic fields observed in the Ap/Bp stars are relics of an earlier evolutionary stage, either generated by dynamo processes during the pre-main sequence stage, or retained from the parental molecular cloud. This hypothesis requires that the field must subsist throughout the various stages of stellar formation without being regenerated. We should therefore expect to observe magnetic fields in some pre-main sequence (PMS) stars of intermediate mass; these are the Herbig Ae/Be stars. We furthermore expect the magnetic fields of these stars should have the same structure and a compatible intensity to the magnetic fields of the Ap/Bp stars. 

In this context, we carried out a survey of many Herbig Ae/Be stars using the new spectropolarimeter ESPaDOnS at the Canada-France-Hawaii Telescope. We observed about 50 stars, clearly detecting a magnetic field in 4 of them (Wade et al. 2005), providing a strong argument in favour of the fossil field hypothesis. We present in this poster the monitoring of two magnetic Herbig stars, HD200775 and V380 Ori, as well as the method employed to determine their magnetic topology.

\section{Observations and data reduction}

Our data were obtained using the high resolution spectropolarimeter ESPaDOnS installed at the CFHT (Canada-France-Hawaii Telescope) (Donati et al. 2006, in preparation), during several observing runs in 2005 and 2006.

We used the ESPaDOnS instrument in circular polarimetric mode, generating Stokes $V$ spectra of 65000 resolution. Each exposure were divided into 4 sub-exposures of equal time in order to compute the optimal extraction of the polarisation spectra (Donati et al. 1997, Donati et al. 2006, in prep.). The data were reduced using the "Libre ESpRIT" package espacially developed for ESPaDOnS (Donati et al. 1997, Donati et al. 2006, in prep.).

\begin{figure}[t]
\begin{minipage}[t]{7.5cm}
\centering
\includegraphics[width=6cm]{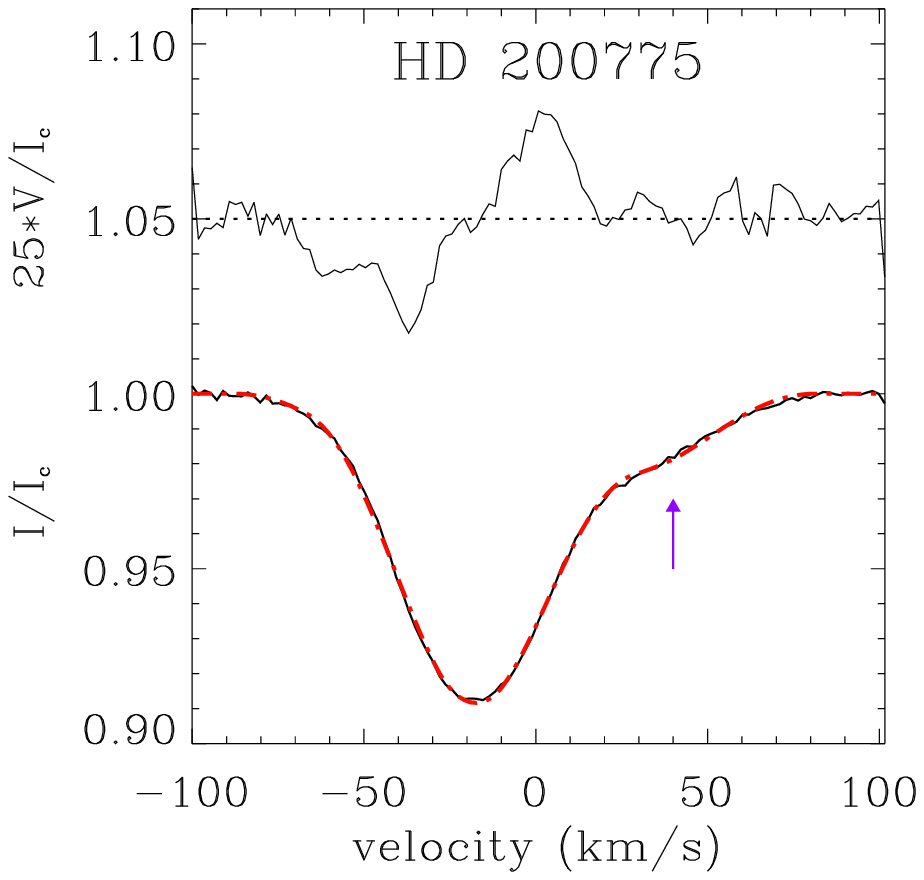}
\end{minipage}\hfill
\begin{minipage}[t]{7.5cm}
\centering
\includegraphics[width=6cm]{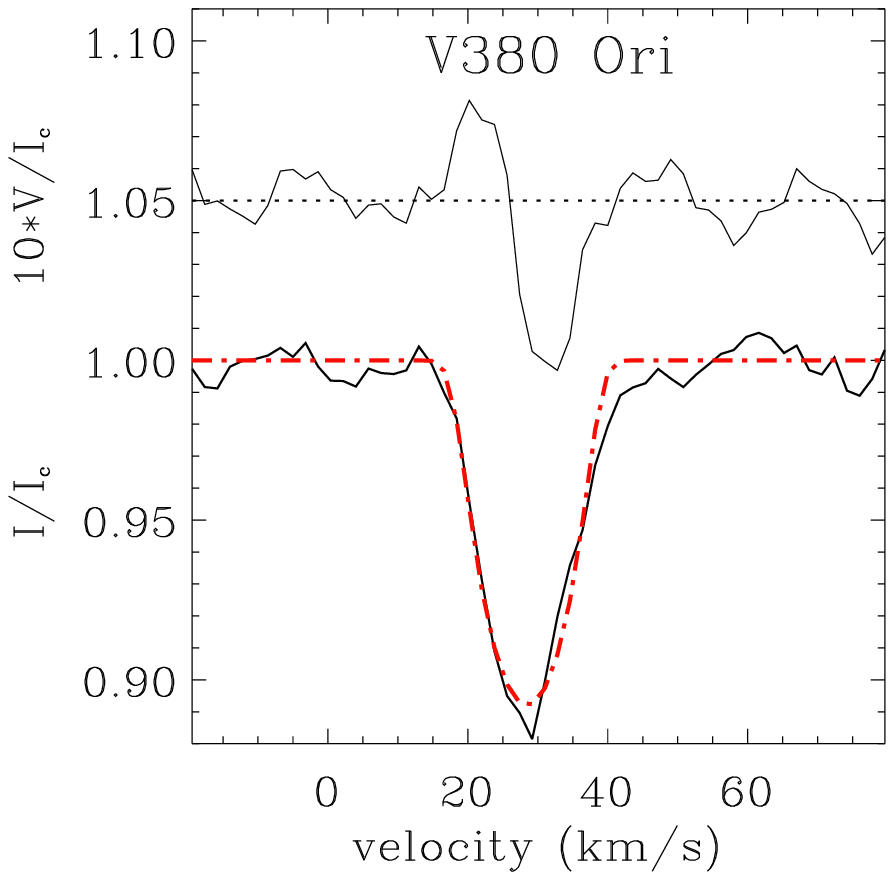}
\end{minipage}
\caption{LSD Stokes $I$ and $V$ profiles of both stars in full line (note the amplification factor in $V$). The Stokes $I$ profile shows that HD 200775 is an SB2 system. The (blue) arrow points to the 2$^{\rm nd}$ component of the system. The dashed (red) lines are the fitted profiles.}
\label{lsdresult}
\end{figure}

We then applied the Least Square Deconvolution (LSD) procedure to all Stokes $I$ and $V$ spectra (Donati et al. 1997), with masks computed using ATLAS 9 models (Kurucz 1993) with effective temperature and surface gravity suitable for these stars (Table \ref{fundparam}). We excluded from this mask hydrogen Balmer lines, strong resonance lines and lines whose Land\'e factor is unknown. Then we cleaned the mask, keeping lines with depth between 0.1 and 0.4 in order to eliminate lines contamined by emission, and adjusted the line depths in order to take into account the relative depth between lines of the observed spectrum. Fig. \ref{lsdresult} shows the resulting LSD Stokes $I$ and $V$ profiles for the two stars. In both cases we see a clear Zeeman signature indicating the presence of a magnetic field in the stellar photosphere.

The Stokes $I$ profile of HD 200775 shows a second component which appears to be due to a companion star with temperature and luminosity comparable to the primary (Alecian et al. 2006, in prep.). Indeed, the secondary's line is visible in most spectra, and it clearly changes position systematically with respect to the primary line. Furthermore, the movement of the two lines in velocity space is consistent with a binary system. However, as the centroid of the Zeeman signature in the $V$ profile is the same that the centroid of the primary component of the $I$ profile, we attribute the detected magnetic field to the primary star of the system (HD 200775A).

\section{Fundamental parameters}

The effective temperatures and luminosities of both stars have been found in the literature, and the masses and radii have been determined from evolutionary models calculated using the CESAM code (Morel 1997). The values of $v\sin i$ have been determined by fitting the photospheric $I$ profiles with the convolution of a (double in the case of HD 200775) gaussian and a rotation profile (Gray 1992). Fig. \ref{lsdresult} shows the fitted profiles superimposed on the observed profiles. Table \ref{fundparam} gives the fundamental parameters of both magnetic stars.

\begin{table}
\caption{Fundamental parameters of HD 200775A and V380 Ori. 1: Hern\'andez et al. (2004), 2: Hamaguchi et al. (2005)}
\label{fundparam}
\begin{center}
\begin{tabular}{llllll}\hline
Star & $T_{\rm eff}$ (K) & $\log(L/L_{\odot})$ & $M/M_{\odot}$ & $R/R_{\odot}$ & $v\sin i$ (km.s$^{-1}$) \\
\hline
HD 200775A & 18600 $\pm$ 2000$^1$ & 3.87 $\pm$ 0.27$^1$ & 10 $\pm$ 2 & 8 $\pm$ 3 & 28.2 $\pm$ 0.2\\
V380 Ori & 10700 $\pm$ 1000$^2$ & 1.9 $\pm$ 0.1$^2$ & 2.8 $\pm$ 0.5 & 2.6 $\pm$ 0.6 & 9.8 $\pm$ 1\\
\hline
\end{tabular}
\end{center}
\end{table}

\section{Fitting of Stokes $V$ profiles}

\begin{figure}[t]
\centering
\includegraphics[angle=90,width=12cm]{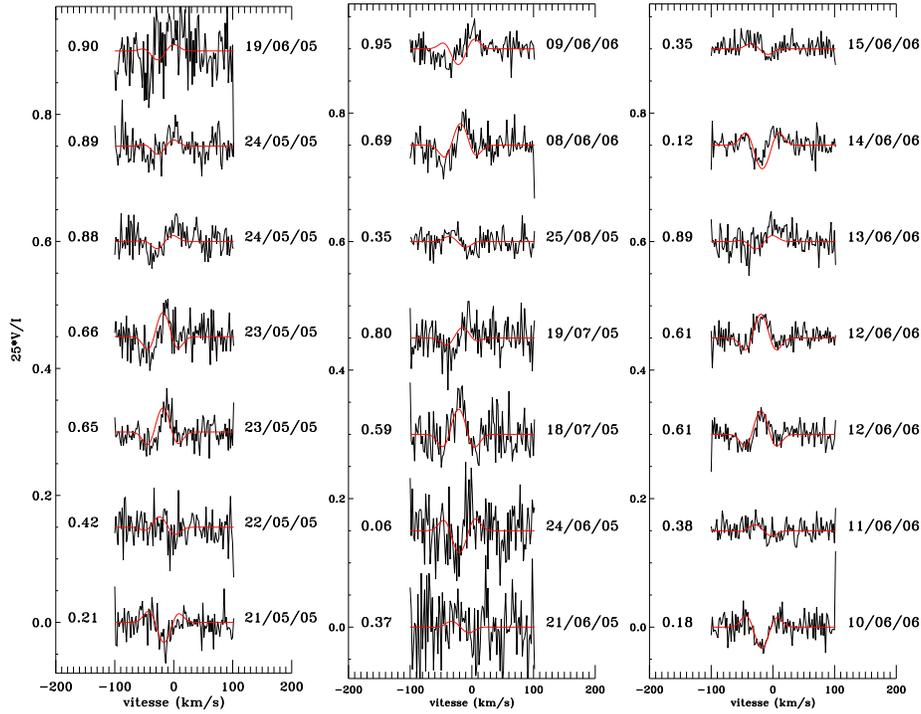}
\caption{LSD $V$ profiles for the 21 spectra of HD 200775A obtained with ESPaDOnS superimposed to the best model.}
\label{plotallv_hd200775}
\end{figure}

For this first attempt at modeling the magnetic fields of HAeBe stars, we calculated an oblique rotator model considering a centered dipole, as described by Stift (1975). We assume that the $I$ profiles are gaussian and we calculated Stokes $V$ profiles using the weak-field Stokes $V$ expression (Landi degl'Innocenti 1973) :
\begin{equation}
	V=-C\overline{g}\lambda_0 cB_{\ell}\frac{dI}{dv}
\end{equation}
where $C=\frac{e}{4\pi mc^2}=4.67\times10^{-13}$\AA$^{-1}$, $\overline{g}$ and $\lambda_0$ are the mean Land\'e factor and wavelength of the lines used in the mask (Sect. 2).
We calculated a grid of $V$ profiles, varying the four free parameters of the model : $\Phi_0$ the initial phase of the ephemeris, $P$ the rotation period of the star, $\beta$ the angle between the rotation and magnetic axes, and $B_{\rm P}$ the dipole magnetic field intensity at pole. Then we fitted simultaneously all Stokes $V$ profiles for each star, selecting models for which the reduced $\chi^2$ was minimised. For the best-fit model of HD 200775A we obtained : $P=4.3$ d, $\beta=90^{\circ}$, $B_{\rm P}=400$ G and $i=\sin^{-1}\left(\frac{P\times v\sin i}{2\pi R}\right)=17^{\circ}$. Fig. \ref{plotallv_hd200775} shows the 21 observed profiles superimposed on the calculated profiles correpsonding to the best model. The discrepancy between the model and the observations at phase 0.05 shows that the model cannot reproduce the observations within the uncertainties. This may be due to to contributions of the secondary, structures on the stellar surface, or multipolar components of the magnetic field not taken into account in our models (Alecian et al. 2006, in preparation). However, it is clear that an oblique rotator dipole model is able to generally reproduce the intensities, morphologies and variations of the Stokes $V$ profiles of HD 200775A.

We have also measured the mean longitudinal magnetic field $B_{\ell}$ from the LSD Stokes profiles of HD 200775. A sinusoidal fit of the temporal variations of $B_{\ell}$ gives a rotation period of 4.38 days, consistent with the period found above.

\begin{figure}[t]
\centering
\includegraphics[angle=90,width=12cm]{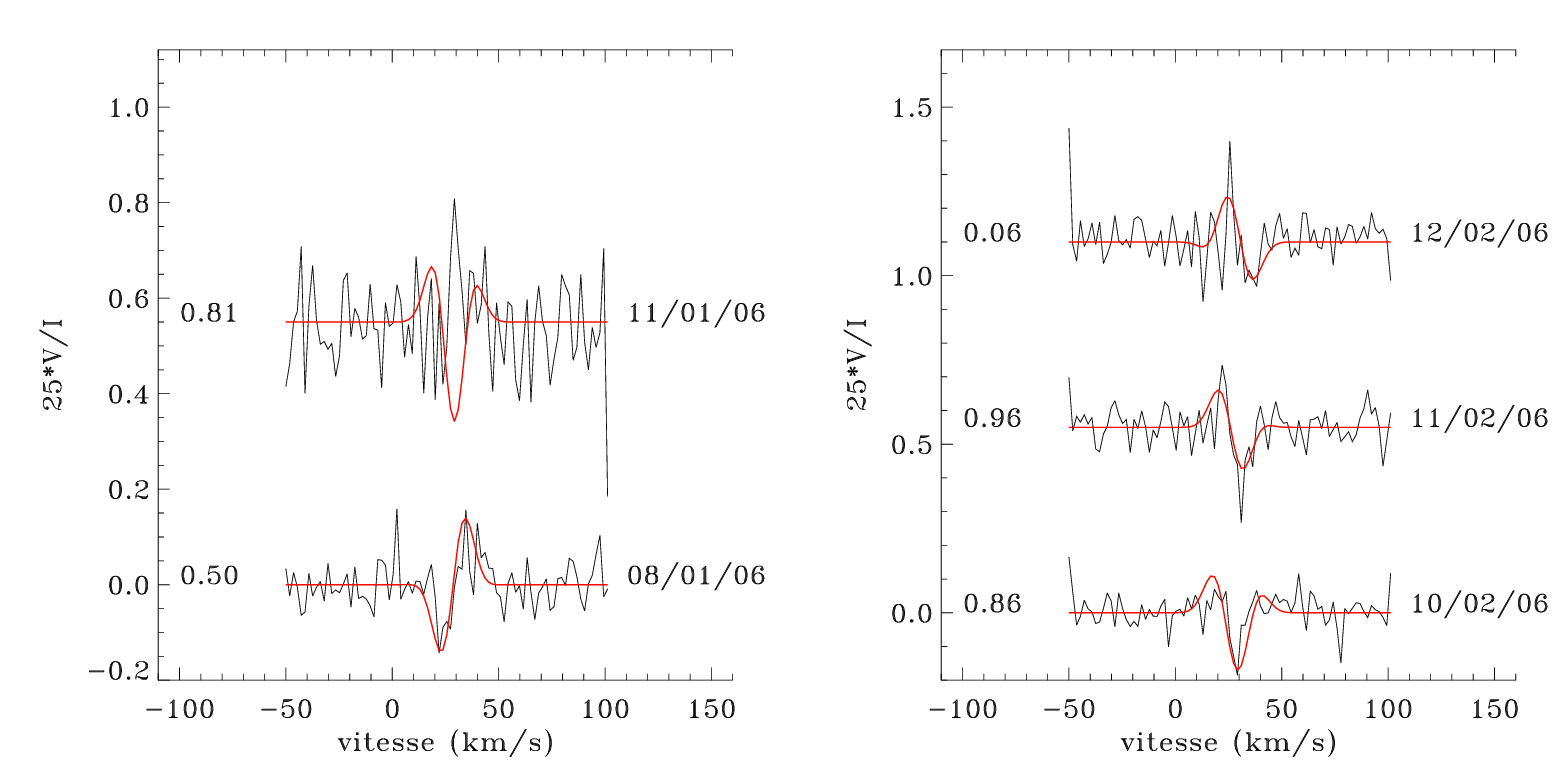}
\caption{Same as Fig \ref{plotallv_hd200775} fo V380 Ori.}
\label{plotallv_v380ori}
\end{figure}

In the case of V380 Ori, because few data were obtained, we obtained two very preliminary solutions. The first one gives : $P=9.8$ d, $\beta=85^{\circ}$, $B_{\rm P}=1400$ G and $i=47^{\circ}$, and the second one gives : $P=7.6$ d, $\beta=90^{\circ}$, $B_{\rm P}=1400$ G and $i=35^{\circ}$. Fig. \ref{plotallv_v380ori} shows the result of the fitting procedure for the first solution. We didn't take into account the February 11th profile in the fitting procedure because its intensity spectrum shows an abnormal increase of the emission in its spectrum, compared to the spectrum observed at other dates (Alecian et al. 2006, in preparation).

\section{Discussion}

Using a simple dipole oblique rotator model we are able to generally reproduce the intensities, morphologies and temporal variations of LSD Stokes $V$ profiles of two magnetic Herbig Ae/Be stars. We find that the dipole intensities of comparable to those observed in the main sequence Ap/Bp stars. 

We have also discovered a magnetic field in the Herbig Ae star HD 190073 (Catala et al. 2006), whose bipolar Stokes $V$ signature does not vary on a timescale of 1 year. This suggests that it hosts a fossil-type magnetic field, and that it either rotates very slowly or is viewed nearly pole-on. We therefore bring an additional argument in support of the fossil field hypothesis for the origin of magnetic fields in intermediate-mass stars: a small fraction of Herbig Ae/Be stars host large-scale, ordered magnetic fields similar to those of the Ap/Bp stars.

\begin{acknowledgements}
\end{acknowledgements}

%\verb|\araa|        & Annual Review of Astronomy and Astrophysics\\
%\verb|\aj|          & Astronomical Journal\\
%\verb|\azh|         & Astronomicheskij Zhurnal\\
%\verb|\aaa|         & Astronomy and Astrophysics\\
%\verb|\aas|         & Astronomy and Astrophysics Supplement Series\\
%\verb|\aar|         & Astronomy and Astrophysics Review\\
%\verb|\apj|         & Astrophysical Journal\\
%\verb|\apjs|        & Astrophysical Journal Supplement Series\\
%\verb|\apss|        & Astrophysics and Space Science\\
%\verb|\baas|        & Bulletin of the American Astronomical Society\\
%\verb|\bsao|        & Bulletin of the Special Astrophysical Observatory\\
%\verb|\ibvs|        & Inform. Bul. Var. Stars\\
%\verb|\jaa|         & Journal of Astronomy and Astrophysics\\
%\verb|\mnras|       & Monthly Notices of the Royal Astronomical Society\\
%\verb|\pasj|        & Publ. Astr. Soc. Japan\\
%\verb|\pasp|        & Publ. of the Astronomical Society of the Pacific\\
%\verb|\pazh|        & Pis'ma v Astronomicheskij Zhurnal\\
%\verb|\sovast|      & Soviet Astronomy\\
%\verb|\sca|         & Scientific American\\
%\verb|\skytel|      & Sky and Telescope\\ 
%\verb|\spsrev|      & Space Science Reviews\\

\end{document}